\documentclass{astrobul}

\usepackage{graphicx}
\usepackage{natbib}

\usepackage{color}

\begin{document}

\journalinfo{2025}{80}{3}{1}[20]

\title{HERCULES STREAM AND THE OUTER ELLIPTICAL RING $R_1R_2$}

\author{A.~M. Melnik\address{1}\email{anna@sai.msu.ru},
  E.~N.~Podzolkova\address{1,2},
  \addresstext{1}{Sternberg Astronomical Institute, Lomonosov Moscow
State University, Universitetskij pr. 13, Moscow 119991,  Russia}
  \addresstext{2}{Faculty of Physics, Lomonosov Moscow State University, Leninskie
Gory 1-2, Moscow 119991, Russia} }

\shortauthor{MELNIK and PODZOLKOVA}

\shorttitle{HERCULES STREAM AND THE RING $R_1R_2$}

\submitted{February 10, 2025; in final form, March 23, 2025}

\begin{abstract}
We study the formation of the Hercules stream in the model Galactic
disk which includes the outer resonance ring $R_1R_2$ located near
the Outer Lindblad Resonance (OLR) of the bar. The Hercules region
and the anti-Hercules region introduced for calibration were
restricted in space by the solar neighborhood, $r<0.5$ kpc, and on
the ($V_R$, $V_T$) plane  by ellipses centered at $V_R=25$ and
$V_T=200$ km s$^{-1}$ (Hercules), and at $V_R=-25$ and $V_T=200$ km
s$^{-1}$ (anti-Hercules). The number of stars in the Hercules region
reaches a maximum at the time period of 2--3 Gyr  from the start of
simulation and the number of stars in the anti-Hercules region
oscillates with a period of $1.8\pm0.1$ Gyr. The majority of stars in
the model disk located in the Hercules and anti-Hercules regions have
orbits elongated perpendicular and parallel to the bar, respectively.
The median value of the initial distances of stars in the Hercules
(anti-Hercules) region is slightly smaller (larger) than the OLR
radius, respectively. There are two types of orbits in the Hercules
region. Orbits of the first type always lie inside a figure bounded
by two ellipses elongated perpendicular to the bar. Orbits of the
second type are elongated at the angles of $ -60$ or $ 60^\circ$ to
the major axis of the bar  most of the time. The distribution of
stars in the Hercules region along the period of slow oscillations in
the angular momentum has two maxima: $P=0.7$ and 2.6 Gyr
corresponding to orbits of the first and second type. In the
anti-Hercules region, most orbits are captured by libration relative
to the major axis of the bar with a period of 1.9 Gyr. In general,
orbits in the Hercules and anti-Hercules regions support the outer
rings $R_1$ and $R_2$ elongated perpendicular and parallel to the
bar, respectively. Stars from the Gaia DR3 catalog located in the
Hercules region appear to be, on average, brighter, bluer and more
luminous  than stars in the anti-Hercules region which is probably
caused by selection effects due to different distributions of these
stellar samples over the Galactic latitude $b$.

\end{abstract}

\section{1. Introduction}

The Hercules stream or the Hercules moving group is a group of stars
located near the Sun that move with a high radial velocity away from
the Galactic center  and with a conspicuously smaller velocity than
that of the rotation-curve velocity in the azimuthal direction. The
Hercules Stream is often identified on the velocity plane ($U$, $V$),
where the positive radial velocity $U$ is directed toward the
Galactic center, and the positive azimuthal velocity $V$ is in the
sense of the Galactic rotation. The position of the Hercules Stream
on the ($U$, $V$) plane  depends on the accepted values of the
velocity components of the solar motion to the apex, but it is often
identified as a moving group with a center located near the values
($-40$, $V_c-40$) km s$^{-1}$ on the ($U$, $V$) plane, where $V_c$ is
the velocity of the rotation curve at the solar distance
\citep{dehnen1998}. The Hercules Stream is named after the star
$\zeta$ Herculis which moves with a similar velocity
\citep{eggen1958}. In general, the definition of the Hercules Stream
remains quite vague.

Hereinafter  we will use radial and azimuthal velocities, $V_R$ and
$V_T$, calculated relative to the Galactic center and corrected for
the solar motion toward the apex, where a positive velocity $V_R$ is
directed away from the Galactic center and a positive velocity $V_T$
is  in the sense of the Galactic rotation.

Modelling of stellar motions in barred galaxies has shown that moving
groups similar to the Hercules Stream arise near the Outer Lindblad
Resonance (OLR) of the bar. The position of the OLR  near the solar
circle gives a constraint  on the value of the angular velocity of
the bar rotation, $\Omega_b$, which must be $\Omega_b \approx 1.9\,
\Omega_0$, where $\Omega_0$ is the angular velocity of the
Galactic-disk rotation  at the solar distance \citep{dehnen2000,
fux2001, minchev2007, antoja2014, monari2017, hunt2018}.

Alternative conceptions are that the Hercules Stream is caused by the
presence of spiral arms in the Galaxy \citep{michtchenko2018}, or by
the motion of stars in the banana-shaped orbits near the Corotation
Radius (CR) of the bar \citep{perez-villegas2017}, or by a collective
perturbation caused by the bar and the spiral arms
\citep{chakrabarty2008, hattori2019}.

\citet{bensby2007} found that stars in the Hercules Stream have a
large range of ages and metallicities. Modern studies with the use of
the LAMOST, APOGEE, and GALAH data have shown that the metallicity of
the Hercules-Stream stars is, on average, slightly higher than the
metallicity of the stars in the surrounding disk \citep{quillen2018,
liang2023}.

The use of the  Gaia  data \citep{prusti2016, brown2021,
vallenari2023} has provided  an opportunity to divide finer structure
within the Hercules stream \citep{ramos2018, lucchini2023}.

We study the formation of the Hercules stream using the model of the
Galaxy in which the disk forms an outer resonance ring $R_1R_2$
located near the OLR of the bar \citep{schwarz1981, buta1991,
byrd1994, buta1995, buta1996, rautiainen1999, rautiainen2000,
melnikrautiainen2009, rautiainen2010, melnik2019}. The elliptical
ring $R_1$ lies slightly closer to the Galactic center and is
stretched perpendicular to the bar while the elliptical ring $R_2$ is
located slightly farther away from the Galactic center and is
elongated parallel to the bar. The backbone of the resonance rings
are periodic orbits near which there are a large number of
quasi-periodic orbits \citep{contopoulos1980, contopoulos1989}.

\citet{weinberg1994} showed that near the Lindblad resonances of the
bar there are orbits that periodically change the direction of their
elongation relative to the  bar major axis. Besides, the angular
momentum and the total energy of  stars change with the same period.
We found that librations of the direction of orbit elongation cause
periodic changes in the morphology of the outer rings and the
appearance of humps on the profiles of the  $V_R$-velocity
distribution along the distance $R$ \citep{melnik2023, melnik2024}.
Orbits captured by resonances of the bar can also create other
kinematical features \citep[][]{fragkoudi2019, chiba2021, trick2021,
drimmel2023}.

\section{2. Model}

We use a 2D  model of the Galaxy with an analytical bar
\citep{freeman1972, athanassoula1983, pfenniger1984} which reproduces
best the distributions of the radial, $V_R$, and azimuthal, $V_T$,
velocities along the Galactocentric distance, $R$, derived from the
Gaia EDR3 and Gaia DR3 data \citep{melnik2021}. The model includes
the bar, exponential disk, classical bulge and halo. The masses of
the bar, disk and bulge are $1.2 \times 10^{10}$, $3.25 \times
10^{10}$ and $5 \times 10^{9}$ M$_\odot$, respectively. The model has
a flat rotation curve on the periphery, and the angular velocity of
the model-disk rotation at the distance of the Sun is $\Omega_0=30$
km s$^{-1}$ kpc$^{-1}$. The bar rotates with the angular velocity of
$\Omega_b=55$ km s$^{-1}$ kpc$^{-1}$ which corresponds to the
positions of the CR and OLR of the bar at $R_{RC}=4.04$ and
$R_{OLR}=7.00$ kpc. The resonance  $-4/1$ of the bar lies at a
distance of $R_{-4/1}=5.52$ kpc. The strength of the bar (the maximum
ratio of the amplitude of the tangential force to the average radial
force at the same radius) is $Q_b=0.3142$ which allows us to classify
it as a moderately strong bar \citep{block2001, buta2004,
diaz-garcia2016}. The bar turns on gradually, gaining its full
strength in 4 bar rotation periods, which corresponds to a growth
time of $T_g=0.45$ Gyr. The model contains $10^{7}$ particles. The
simulation time is 6 Gyr.

The best agreement between the model and observations corresponds to
the position angle of the Sun relative to the major axis of the bar
of $\theta_\odot=-45^\circ$. Since our model has the  order of
symmetry $m=2$, both values of the position angle,
$\theta_\odot=-45^\circ$ and $135^\circ$, are equivalent \citep[for
more details see][]{melnik2021}.

The time intervals at which the model and observations  agree best
correspond to the periods $t=2.5\pm0.3$ and $4.5\pm0.5$ Gyr from the
start of  simulation, when the humps on the profiles of the
$V_R$-velocity distribution  disappear \citep{melnik2024}.
Hereinafter, we will use the time period of $t=2.5$--2.6 Gyr to
compare the model and observations.

\section{3. Observations}

\begin{figure*}
\centering  \resizebox{11 cm}{!}{\includegraphics{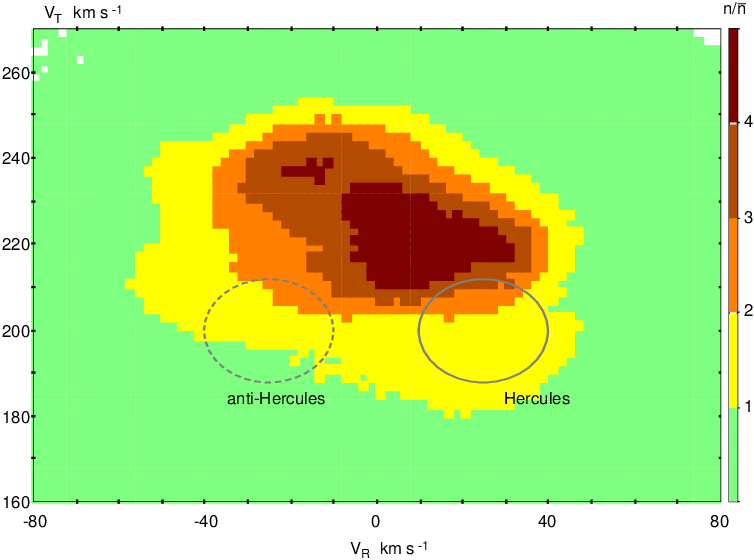}}
\caption{Distribution of Gaia DR3 stars on the  ($V_R$, $V_T$) plane.
The color shows the ratio of the number of stars in $2\times 2$ km
s$^{-1}$  areas to the average number of stars in the areas
containing at least 10 stars (see the color palette on the right). We
consider stars located at the distance $r<0.5$ kpc from the Sun, near
the Galactic plane, $|z|<200$ pc, with the parallax to parallax error
ratio $\varpi/\varepsilon_\varpi>5$, the error of
$\textrm{RUWE}<1.4$, the vertical velocity $|V_z|< 50$ km s$^{-1}$,
and the line-of-sight velocity measured by the Gaia spectrometer. The
positions of the Hercules and anti-Hercules regions are shown by
ellipses centered at $V_R=25$ and $V_T=200$ km s$^{-1}$(Hercules,
solid line), and at $V_R=-25$ and $V_T=200$ km s$^{-1}$
(anti-Hercules, dotted line), with semi-axes of $\Delta V_R=15$ and
$\Delta V_T=12$ km s$^{-1}$.  } \label{fig_gaia}
\end{figure*}

We consider stars from the Gaia DR3 catalog \citep{prusti2016,
katz2018, brown2021, vallenari2023} located at the distance $r<0.5$
kpc from the Sun and near the Galactic plane, $|z|<0.2$ kpc, having
the parallax to parallax error ratio $\varpi/\varepsilon_\varpi>5$,
the error  $\textrm{RUWE}<1.4$ and the line-of-sight velocity $V_r$
measured by the Gaia spectrometer. Fig.~\ref{fig_gaia} shows the
distribution of the observational sample of stars on the ($V_R$,
$V_T$) plane. The color shows the ratio of the number of stars in the
$2\times 2$ km s$^{-1}$ areas  to the average number of stars in the
areas containing at least 10 stars. It is clearly seen that the
observational distribution is elongated towards positive values of
the velocity $V_R$ and low values of the azimuthal velocity $V_T$.

It is quite difficult to  compare directly the model and
observations. First, the observational and model samples have
different numbers of objects. Second, the observational sample
includes stars of the thick disk and halo, which are absent in the
model sample. Third, the observational sample is burdened by
selection effects, the main one of which is related to the selection
of stars with a known line-of-sight velocity $V_r$.

The effective limit on the measurement of  magnitudes $G$ and
astrometry of Gaia is $1\,050\,000$ sources/degree$^2$
\citep{prusti2016}. When the Gaia processing program chooses which of
two stars to assign a measurement window to, it will always choose
the brighter star, so the crowding of stars reduces the completeness
of the catalog due to lack of faint objects. The size of the windows
for measuring BP/RP photometry and line-of-sight velocities is even
larger, so the limits on the number of simultaneously measured
objects are lower here and equal to $750\,000$ and $35\,000$
sources/degree$^2$, respectively. The probability that the Gaia
spectrometer will successfully measure the line-of-sight velocity,
$V_r$, of a star  depends on its magnitude $G_{RVS}$, color $G-
G_{RP}$ and the density of bright stars in a given region of the sky
\citep{boubert2020, rybizki2021, everall2022, castro-ginard2023}.

To calibrate the number of stars, we use the anti-Hercules region
located symmetrically to the Hercules region relative to the line
$V_R=0$. Fig.~\ref{fig_gaia} shows the positions of the Hercules and
anti-Hercules regions which are restricted by ellipses with centers
at $V_R=25$ and $V_T=200$ km s$^{-1}$ (Hercules), and at $V_R=-25$
and $V_T=200$ km s$^{-1}$ (anti-Hercules) and semi-axes of $\Delta
V_R=15$ and $\Delta V_T=12$ km s$^{-1}$.  In addition, we excluded
from consideration stars with a high vertical velocity $V_z$, $|V_z|>
50$ km s$^{-1}$. The velocity dispersion of Gaia DR3 stars in the
vertical direction at the solar distance is $\sigma_z=15.46$ km
s$^{-1}$, therefore the restriction $|V_z|> 50$ km s$^{-1}$
corresponds to $\sim3\sigma$. This resulted in exclusion of 1.3 and
2.2\%  stars from the Hercules and anti-Hercules regions,
respectively. The final samples of Gaia DR3 stars in the Hercules and
anti-Hercules regions include $148\,404$ and $108\,005$ objects,
respectively (see also Section 4.7).

When calculating the  $V_R$, $V_T$ and $V_z$ velocities relative to
the Galactic center \citep[][Eq.~3, 4 and 5]{melnik2021}, we use the
following data for the solar  motion. The components of the solar
motion in the direction toward the Galactic center, in the sense  of
the Galactic rotation and perpendicular to the Galactic plane, as
well as the angular velocity of the disk rotation at the solar
distance  are taken to be  ($U_\odot$, $V_\odot$, $W_\odot$)=(10, 12,
7) km s$^{-1}$ and $\Omega_0=30$ km s$^{-1}$ kpc$^{-1}$,
respectively, which is consistent with the kinematics of OB
associations \citep{melnik2020}.

The Galactocentric distance of the Sun is adopted  to be $R_0=7.5$
kpc \citep[][]{glushkova1998, nikiforov2004, eisenhauer2005,
bica2006, nishiyama2006, feast2008, groenewegen2008, reid2009b,
dambis2013, francis2014, boehle2016, branham2017, iwanek2023}. In
general, the choice of $R_0$ in the range of 7--9 kpc has practically
no effect on our results. With this choice of $R_0$, the azimuthal
rotation velocity of the disk at the solar distance is $V_T=225$ km
s$^{-1}$.

\section{4. Results}

\subsection{4.1 Distribution of model stars on the  ($V_R$, $V_T$) plane at different time periods}

\begin{figure*}
\centering  \resizebox{11 cm}{!}{\includegraphics{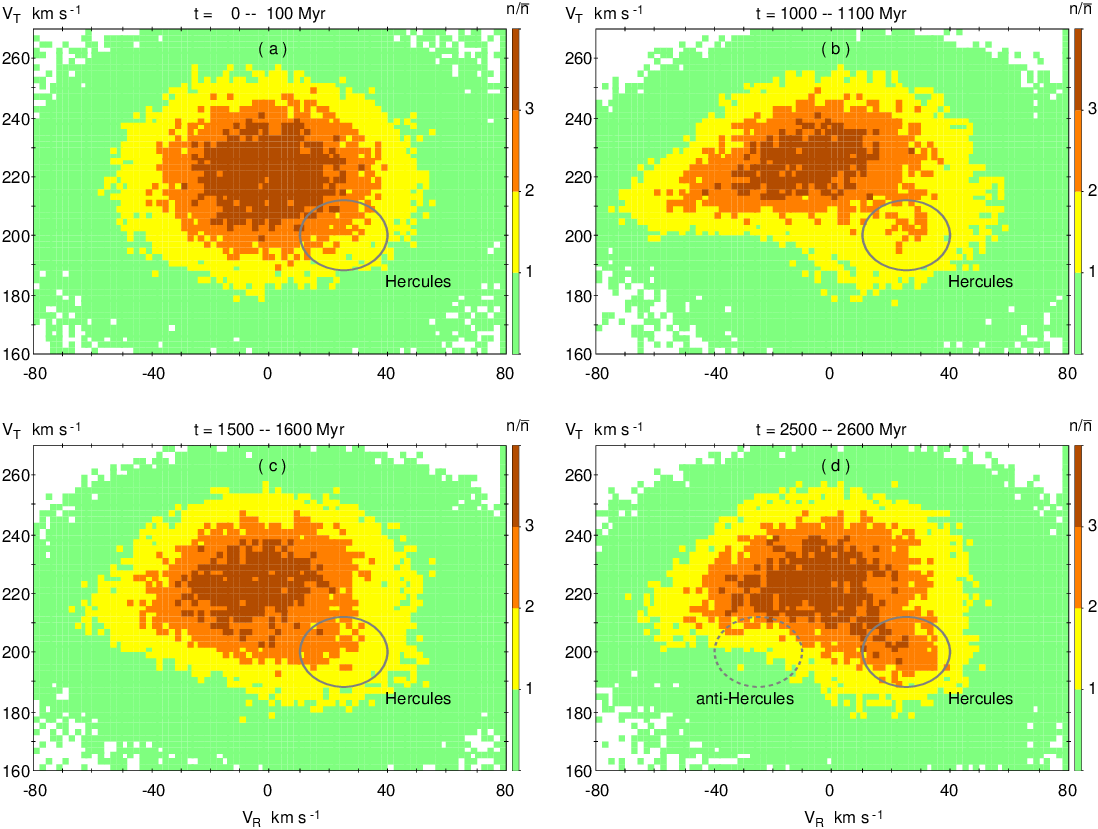}}
\caption{Distribution of model stars located at the distance of
$r<0.5$ kpc from the Sun on the ($V_R$, $V_T$) plane at the different
time periods. We calculated the number of stars $n$ in areas of
$2\times 2$ km s$^{-1}$ in size and compared them with the average
number of stars, $\overline{n}$, in similar areas containing at least
one star, $n \ge 1$, in the region of $-80<V_R<80$ and $160<V_T<270$
km s$^{-1}$. As the ratio $n/\overline{n}$ increases, the color of
the area changes from green to brown (see the color palette on the
right). The position of the Hercules region is shown by the ellipse
with the center at $V_R=25$ and $V_T=200$ km s$^{-1}$ and the
semi-axes of $\Delta V_R=15$ and $\Delta V_T=12$ km s$^{-1}$,
respectively. It is clearly seen that the Hercules region includes
maximum number of stars  at the time period close to $t=2500$ Myr
from the start of simulation. Frame (d) also shows the position of
the anti-Hercules region (dotted line), which is needed for
calibration. } \label{fig_vr_vt_4}
\end{figure*}

The Hercules region  forms rather slowly  in the model disk.
Fig.~\ref{fig_vr_vt_4} shows the distribution of  model stars located
at the distance of $r<0.5$ kpc from the Sun on the  ($V_R$, $V_T$)
plane at the different time periods. We divided the region
$-80<V_R<80$ and $160<V_T<270$ km s$^{-1}$ into  $2\times 2$ km
s$^{-1}$ areas, calculated the number of stars $n$ in each area at
different times and compared them with the average number of stars,
$\overline{n}$, in areas containing at least one star, $n \ge 1$, in
the range $-80<V_R<80$ and $160<V_T<270$ km s$^{-1}$. As the ratio
$n/\overline{n}$ increases, the color of the area changes (see the
color palette on the right).

We consider the time periods of $t=0$--100, 1000--1100, 1500--1600,
and 2500--2600 Myr from the start of simulation. To increase the
number of stars in the model sample, we sum up the number of stars
$n$ that fall into the circle $r<0.5$ kpc at 10 time instants
separated from each other by 10 Myr within   the time interval of 100
Myr. At each of the moments considered, the circle $r<0.5$ kpc, where
$r$ is the heliocentric distance, includes different stars.

The position angle of the Sun relative to the major axis of the bar
is adopted to be $\theta_\odot=135^\circ$. In the reference frame of
the bar, stars at the distance of the Sun rotate in the direction
opposite that of Galactic rotation with the average velocity of
$R_0(\Omega_0-\Omega_b)$, which is 188 km s$^{-1}$. Over 10 Myr, they
move on average by 1.9 kpc, which is significantly larger than the
size of the circle of $r<0.5$. In general, this estimate is obtained
for circular orbits, but real orbits may have loops. However,
verification showed that the total sample does not actually include
the same stars at different time moments.

Fig.~\ref{fig_vr_vt_4} shows that the distribution of stars on the
($V_R$, $V_T$) plane changes over time. At the initial moment, the
distribution of stars has an elliptical shape with the center at
$V_R=0$ and $V_T=\Omega_0 R_0-7=218$ km s$^{-1}$. The difference $-7$
km s$^{-1}$ between the most probable velocity at the initial moment
and the velocity of the rotation curve, $\Omega_0 R_0$, is a
consequence of the asymmetric drift \citep[for
example,][]{binney2008}. We can clearly see that  the distribution of
stars changes with time forming branches located both to the right
and to the left from the central condensation. The so-called Hercules
stream corresponds to the positive radial velocities of 20--30 km
s$^{-1}$ and  azimuthal velocities of $V_T\approx200$ km s$^{-1}$,
lagging behind the rotation curve velocity ($V_c=225$ km s$^{-1}$) by
$\sim 25$ km s$^{-1}$. For clarity, we define the Hercules stream as
the region within an ellipse centered at $V_R=25$ and $V_T=200$ km
s$^{-1}$ and semi-axes of $\Delta V_R=15$ and $\Delta V_T=12$ km
s$^{-1}$, respectively. It is evident that the Hercules region
includes the greatest number of model stars at the time period
$t=2.5$--2.6 Gyr from the start of simulation
(Fig.~\ref{fig_vr_vt_4}d).

Hereinafter, we  will  also consider the anti-Hercules region which
is needed for  comparison between the model and observations. The
Hercules and anti-Hercules regions are located symmetrically relative
to the vertical axis $V_R=0$. The strict definition of the Hercules
and anti-Hercules regions/streams is as follows: a star belongs to
the Hercules or anti-Hercules region/stream if it passes within 0.5
kpc of the Sun during the period considered and its velocities $V_R$
and $V_T$ lie within  the ellipse centered at $V_R=25$ and $V_T=200$
km s$^{-1}$ (Hercules), or at $V_R=-25$ and $V_T=200$
(anti-Hercules), with semi-axes $\Delta V_R=15$ and $\Delta V_T=12$
km s$^{-1}$ (Fig.~ \ref{fig_vr_vt_4}d).

\subsection{4.2 Variations in the number of stars in the Hercules and anti-Hercules regions over time}

\begin{figure*}
\centering  \resizebox{11 cm}{!}{\includegraphics{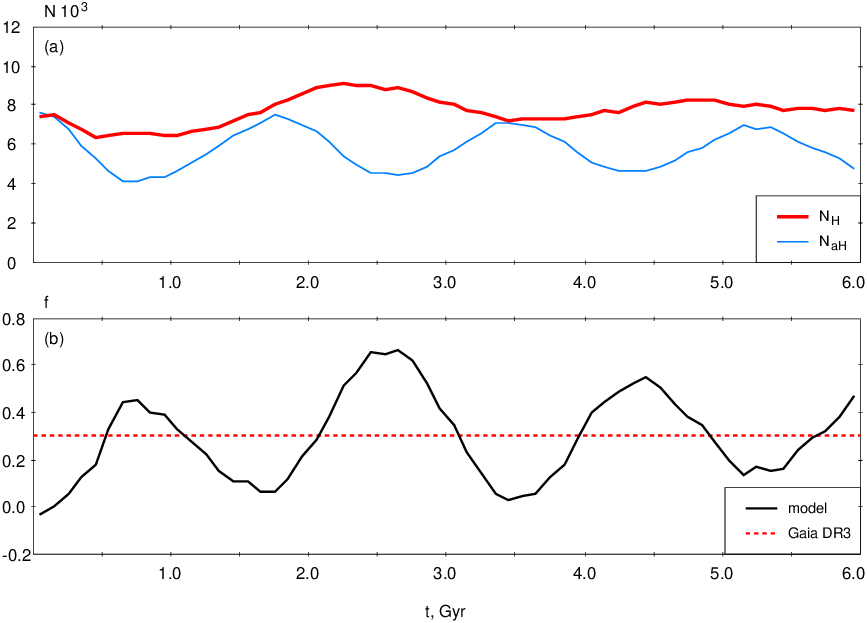}}
\caption{(a) Variations in the number of model stars in the Hercules
region ($N_H$, red line) and in the anti-Hercules region ($N_{aH}$,
blue line). (b) Variations in the relative difference, $f$, of stars
in the Hercules and anti-Hercules regions (Eq.~\ref{frac}) over time
calculated for model stars (black curve) and for  Gaia DR3 stars
($f_g=0.315$, red dashed line).} \label{fig_time_n}
\end{figure*}

We calculated the numbers of model stars in the Hercules, $N_H$, and
anti-Hercules, $N_{aH}$, regions over the time period 0--6 Gyr. Each
of the 60 values of $N_H$ and $N_{aH}$ represents the sum of the
instantaneous star numbers obtained for 10 moments separated by 10
Myr. Fig.~\ref{fig_time_n}a shows that the star number in the
Hercules region ($N_H$, red line) reaches the first  maximum at the
interval 2.0--2.6 Gyr, then decreases. The second maximum $N_H$
occurs during the period 4.3--5.2 Gyr and looks more like a plateau.
The number of stars in the anti-Hercules region ($N_{aH}$, blue line)
shows well-defined oscillations with a period of $P=1.8\pm0.1$ Gyr.
We previously observed a change in the number of stars in some
regions with the period close to $P\approx 2.0$ Gyr, which is caused
by librations of  orbits near the OLR \citep{melnik2023, melnik2024}.
Note that the maximum in the distribution of $N_H$ (red line) at the
time period $t=2.0$--2.6 Gyr coincides with the minimum in the
distribution of $N_{aH}$ (blue line).

Fig.~\ref{fig_time_n}b shows variations of the relative difference
$f$ in the number of stars  in the Hercules and anti-Hercules
regions:

\begin{equation}
f=2\frac{N_H- N_{aH}}{N_H+ N_{aH}}, \label{frac}
\end{equation}

\noindent which does not depend on the number of stars neither in the
model nor  observational samples.

Fig.~\ref{fig_time_n}b also shows the observational value of $f$
(dashed line) calculated for Gaia DR3 stars located in the region
$r<0.5$ and $|z|<0.2$ kpc and in the ellipses on the  ($V_R$, $V_T$)
plane corresponding to the Hercules and anti-Hercules regions. The
observational value of $f$ is $f_g=0.315\pm0.004$. For  model stars,
minimum and maximum values of $f$ are $-0.009$ and 0.656,
respectively, with an average value of $\overline{f}=0.291\pm0.011$
obtained over the time interval 0--6 Gyr. Thus, the observational
value of $f_g$ calculated for Gaia DR3 stars and the average value of
$f$ obtained for model stars are consistent within $1.6\sigma$. A
comparison of the model and observational values of $f$ shows that
there are many time moments when they coincide.

However, the observational value of $f_g$ might be significantly
underestimated. The sample of Gaia stars in the vicinity of $r<0.5$
kpc from the Sun contains a noticeable number of thick-disk and halo
stars, even inside the layer $|z|<200$ pc. Suppose that the velocity
distribution of thick-disk and halo stars in the region of $r<0.5$
kpc is symmetric relative to the line $V_R=0$. In this case,  the
numerator in Eq.~\ref{frac} would not be affected by selection
effects, since the number of thick-disk and halo stars in the
Hercules and anti-Hercules regions must be approximately the same.
However, the denominator in Eq.~\ref{frac} must include twice the
number of thick-disk and halo stars. Therefore, the value of $f_g$
calculated solely for thin-disk stars can be noticeably higher than
the value of $f_g=0.315$ obtained here. In addition, selection
effects related to the choice of the brightest stars for
line-of-sight velocity measurements in crowded regions also lead to
an underestimation of the $f_g$ value (see Section 4.7).

\subsection{4.3 Orientation of orbits in the Hercules and anti-Hercules regions}

\begin{figure*}
\centering  \resizebox{14 cm}{!}{\includegraphics{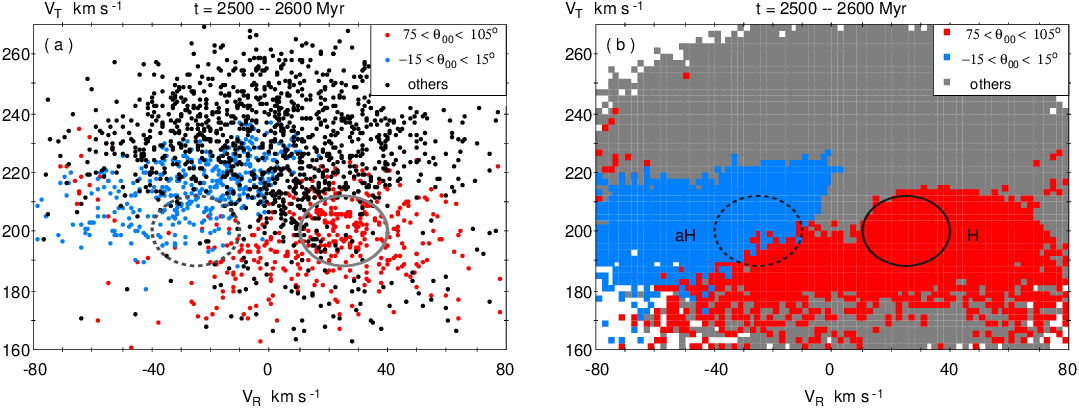}}
\caption{(a) Distribution of model stars on the  ($V_R$, $V_T$) plane
and their orbital orientation relative to the bar. Stars with
elliptical orbits  elongated perpendicular to the bar
($75\le\theta_{00}<105^\circ$) are shown in red; stars with
elliptical orbits ($n_s=2$) elongated parallel to the bar
($0\le\theta_{00}<15^\circ \cup 165\le\theta_{00} < 180^\circ$) are
shown in blue; all other cases are shown in gray. The angle
$\theta_{00}$ determines the average orientation of the orbit
relative to the major axis of the bar at the time  period 0--3 Gyr.
2\% of stars are represented. We consider model stars  that fall in
the neighborhood of the Sun, $r<0.5$ kpc, at the time period of
$t=2.5$--2.6 Gyr. Also are shown the boundaries of the Hercules
(solid line) and anti-Hercules (dashed line) regions. (b)
Distribution of areas of $2\times 2$ km s$^{-1}$ in size on the
($V_R$, $V_T$) plane, the color of which shows the most probable
orientation of the orbits. The areas in which the majority of stars
have elliptical orbits ($n_s=2$) elongated perpendicular to the bar
($75\le\theta_{00}<105^\circ$) are shown in red; the areas in which
most of stars have elliptical orbits ($n_s=2$) elongated parallel to
the bar ($0\le\theta_{00}<15^\circ \cup 165\le\theta_{00} < 180^
\circ$) are shown in blue; areas in which most of  stars have another
orientation of orbits or a different order of  symmetry are shown in
gray. It is clearly seen that the majority of stars in the Hercules
and anti-Hercules regions have elliptical orbits elongated
perpendicular and parallel to the bar, respectively.}
\label{fig_orient}
\end{figure*}

Fig.~\ref{fig_orient}a shows the distribution of  model stars on the
($V_R$, $V_T$) plane. We consider only  stars that fall within the
vicinity of $r<0.5$ kpc from the Sun at the time period $t=2.5$--2.6
Gyr from the start of simulation. We divided model stars  into three
groups depending on  orientation of their orbits: elliptical orbits,
i.e. orbits with the order of symmetry  $n_s=2$, elongated
perpendicular to the bar ($75\le\theta_{00}<105^\circ$); elliptical
orbits ($n_s=2$) elongated parallel to the bar
($0\le\theta_{00}<15^\circ$ or $165\le\theta_{00} < 180^\circ$); and
other orbits. The angle $\theta_{00}$ determines the average
orientation of the orbit relative to the major axis of the bar at the
time period $t=0$--3 Gyr from the start of simulation. To avoid
overloading the Figure with objects, we  show only 2\% of stars. The
boundaries of the Hercules and anti-Hercules regions are also
outlined. It is clearly seen that the majority of stars  in the
Hercules region have orbits elongated perpendicular to the bar while
in the anti-Hercules region orbits are stretched parallel to the bar.

Fig.~\ref{fig_orient}b shows the most probable orientation of orbits
in different regions of the  ($V_R$, $V_T$) plane. We divided the
($V_R$, $V_T$) plane in the range of values $-80<V_R<80$ and
$160<V_T<270$ km s$^{-1}$ into $2\times 2$ km s$^{-1}$ areas.  For
each area, we calculated the number of stars, $n_{\perp}$, whose
orbits have an elliptical shape ($n_s=2$) and are stretched
perpendicular to the bar ($75\le\theta_{00}<105^\circ$); the number
of stars, $n_{\parallel}$, whose orbits have an elliptical shape
($n_s=2$) and are elongated parallel to the bar
($0\le\theta_{00}<15^\circ$ or $165\le\theta_{00} < 180^\circ$); and
the number of stars, $n_{oth}$, with a different value of $n_s$
($n_s\neq 2$) or a different orbital orientation. If most  stars in
the area have elliptical orbits elongated perpendicular to the bar
($n_{\perp}>n_{\parallel}$ and $n_{\perp}>n_{oth}$), the area is
shown in red; if most stars have elliptical orbits elongated parallel
to the bar ($n_{\parallel}>n_{\perp}$ and $n_{\parallel}>n_{oth}$),
then it is shown in blue; and if most  orbits have a different order
of symmetry or a different orientation ($n_{oth}>n_{\perp}$ and
$n_{oth}>n_{\parallel}$), then the area is shown in gray. It is
clearly seen that the majority of stars in the Hercules and
anti-Hercules regions have elliptical orbits elongated perpendicular
and parallel to the bar, respectively.

\begin{figure*}
\centering  \resizebox{14 cm}{!}{\includegraphics{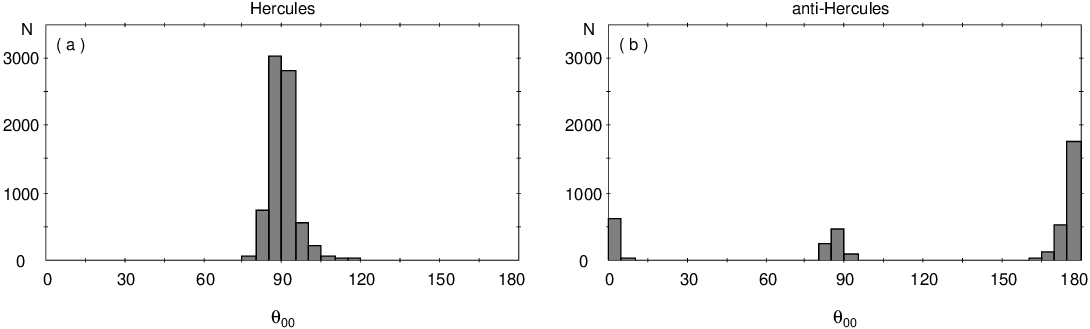}}
\caption{Distributions of elliptical orbits ($n_s=2$) along the angle
$\theta_{00}$ in the (a)  Hercules  and (b) anti-Hercules  regions.
The angle $\theta_{00}$ lies in the range $0\le\theta_{00}<180^\circ$
and determines the average orientation of the orbit relative  to the
major axis of the bar at the time period $t=0$--3 Gyr. We consider
model stars which fall in the solar neighborhood, $r<0.5$ kpc, at the
time period of $t=2.5$--2.6 Gyr. It is clearly seen that most stars
in the Hercules region  have orbits elongated perpendicular to the
bar ($75\le\theta_{00}< 105^\circ$) while those in the anti-Hercules
region have orbits elongated parallel to the bar, specifically at the
angle $\theta_{00}$ in the range $165\le\theta_{00} < 180^\circ$. }
\label{fig_his_theta}
\end{figure*}

Fig.~\ref{fig_his_theta} shows the distributions of elliptical orbits
along the angle $\theta_{00}$ in the Hercules and anti-Hercules
regions. The fraction of elliptical orbits ($n_s=2$) in these regions
reaches 85\% and 87\%, respectively. Fig.~\ref{fig_his_theta} shows
that almost all stars (84\%) in the Hercules region have orbits
elongated perpendicular to the bar ($75\le\theta_0< 105^\circ$). In
contrast, in the anti-Hercules region most stars have orbits
elongated parallel to the bar: 54\% are oriented at the angle
$\theta_{00}$ in the range of $165\le\theta_{00} < 180^\circ$ and
14\% at the angle $\theta_{00}$ in the range of $0\le\theta_{00}\le
15^\circ$; and only 18\% are elongated perpendicular to the bar
($75\le\theta_{00}<105^\circ$).

Among all model stars  that fall in the solar neighborhood, $r<0.5$
kpc, at the time period of $t=2.5$--2.6 Gyr, 21\% have elliptical
orbits ($n_s=2$) elongated perpendicular to the bar
($75\le\theta_{00}<105^\circ$), and 19\% have elliptical orbits
($n_s=2$) elongated parallel to the bar ($0\le\theta_{00}<15^\circ$
or $165\le\theta_0 < 180^\circ$). Moreover, in the latter case, 11\%
have orbits oriented at the angle of $\theta_{00}$ in the range of
$0\le\theta_{00}<15^\circ$ and 8\%  in the range of
$165\le\theta_{00} < 180^\circ$. Thus, the distribution of orbits
along the  angle $\theta_{00}$ in the Hercules and anti-Hercules
regions differs significantly from the general distribution of stars
in the vicinity of the Sun.

\subsection{4.4 Initial values of $R$, $V_T$ and $V_R$ in the Hercules and
anti-Hercules regions}

\begin{figure*}
\centering  \resizebox{14 cm}{!}{\includegraphics{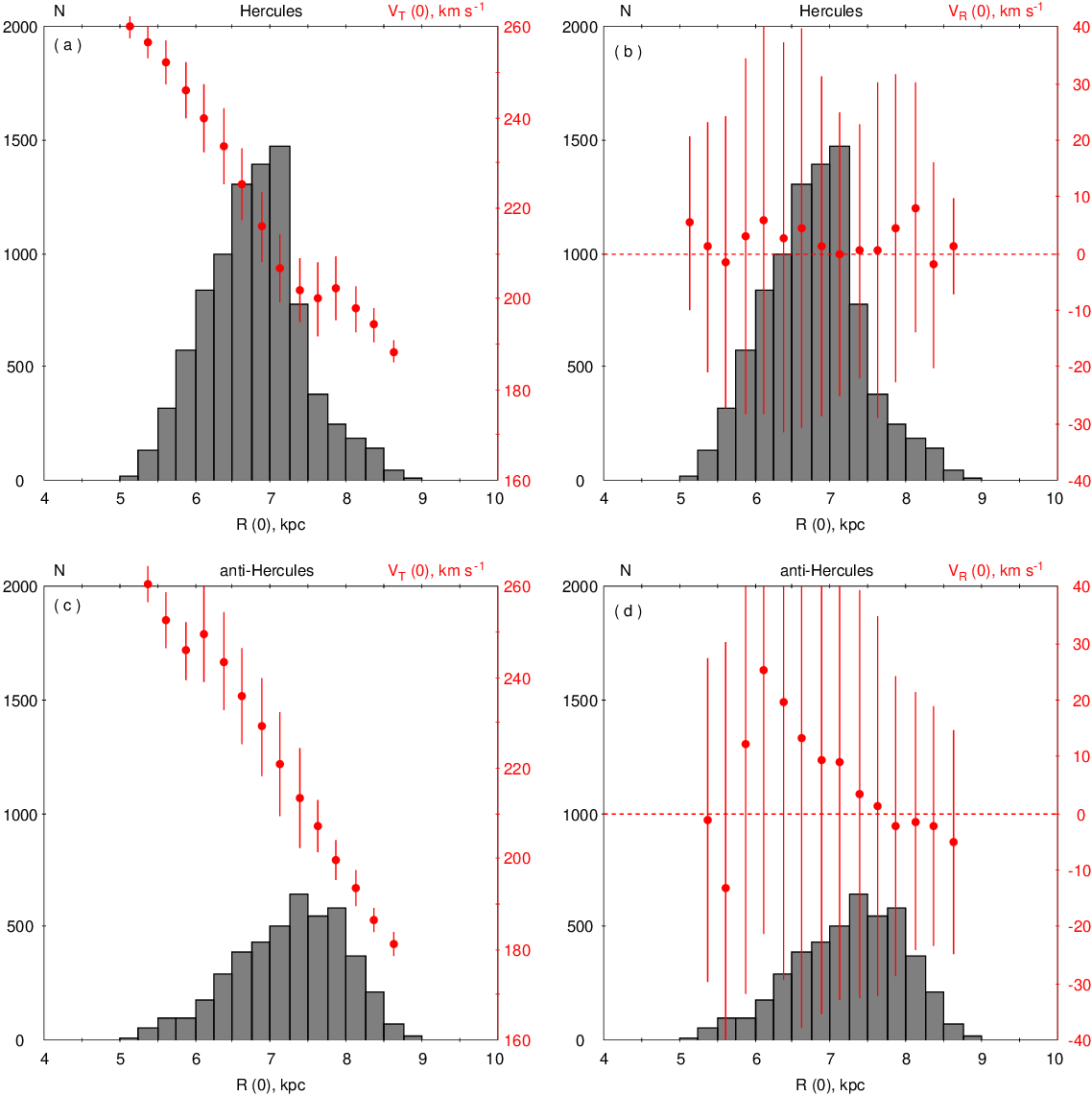}}
\caption{Distributions of the numbers of model stars $N$ along the
initial value of the distance $R(0)$ (gray bars) in the (a, b)
Hercules and (c, d) anti-Hercules regions. Other distributions are
superimposed upon these histograms: (a, c) the median values  of the
initial azimuthal velocity $V_T(0)$ and (b, d) the median values of
the initial radial velocity $V_R(0)$ (red circles) calculated in each
bin along $R$. The red vertical lines show the $+/-$ dispersion of
the median velocities. The scales of  variations in the $V_T$ and
$V_R$ velocities are shown on the right vertical axes. We can clearly
see the general tendency for  the $V_T$-variations: a decrease in the
value of $V_T(0)$ with  increasing $R(0)$, although there is a small
plateau (7.0--8.0 kpc) in the Hercules region. The median radial
velocity $V_R(0)$ in the Hercules region fluctuates in the range
$[-2, 8]$ km s$^{-1}$ while in the anti-Hercules region the velocity
$V_R(0)$ sharply decreases from 25 to 3 km s$^{-1}$ at the interval
6.0--7.5 kpc and  fluctuates in the range  $[-5, 3]$ km s$^{-1}$ at
the interval 7.5--8.5 kpc. In general, the bins along $R$ containing
maximum number of stars correspond to nearly zero value of the
initial radial velocity, $V_R(0)$. We consider model stars which fall
in the solar vicinity, $r<0.5$ kpc, at the time period $t=2.5$--2.6
Gyr.} \label{fig_his_r0}
\end{figure*}

Fig.~\ref{fig_his_r0} shows the distributions of the numbers of model
stars $N$ in the Hercules and anti-Hercules regions along the initial
value of the Galactocentric distance $R(0)$. The bin width along
$R(0)$ is 250 pc. We consider  model stars that fall in the solar
neighborhood, $r<0.5$ kpc, at the time period $t=2.5$--2.6 Gyr. The
median values of the initial distances $R$ in the Hercules and
anti-Hercules regions are 6.8 and 7.3 kpc, respectively, i.~e. most
stars arrive in the Hercules (anti-Hercules) region from distances
smaller (larger) than the OLR radius ($R_{OLR}=7.0$ kpc).

Fig.~\ref{fig_his_r0}a and Fig.~\ref{fig_his_r0}c also show the
median values of the initial azimuthal velocity $V_T(0)$ and the
$+/-$ dispersion in $V_T$ calculated in each bin along $R(0)$. The
general tendency is clearly visible: a decrease in the value of $V_T$
with increasing $R$, although there is a small plateau (7.0--8.0 kpc)
in the Hercules region. The decrease in the initial value of $V_T$
with increasing $R$ may be due to highly elongated orbits. In order a
star to reach quickly the OLR radius  from the initial distance $R$
less (greater) than $R_{OLR}=7.0$ kpc, it must have  a greater
(lower) azimuthal velocity than stars have, on average, at this
distance.

Fig.~\ref{fig_his_r0}b and Fig.~\ref{fig_his_r0}d show the median
values of the initial radial velocity $V_R(0)$ and the $+/-$
dispersion in $V_R$. The median radial velocities $V_R(0)$ in the
Hercules region vary in the range $[-2, 8]$ km s$^{-1}$.
Interestingly, the median radial velocity in the bin
$R(0)=7.00\textrm{--}7.25$ kpc, corresponding to  maximum number of
stars $N$  is $V_R(0)\approx 0$. This is due to the fact that the
most probable value of the initial radial velocity is zero,
$V_R(0)=0$, so there are a lot of such stars in the model.

As for the anti-Hercules region (Fig.~\ref{fig_his_r0}d), here the
median value of the  initial velocity $V_R(0)$ sharply decreases from
25 to 3 km s$^{-1}$ at the interval 6.0--7.5 kpc and  fluctuates in
the range $[-5, 3 ]$ km s$^{-1}$ at the interval 7.5--8.5 kpc. Once
again, we observe that the bins with  maximum number of particles $N$
correspond to $V_R(0)\approx 0$.

The dispersion of the initial  azimuthal velocities in the Hercules
and anti-Hercules regions is $\sigma_T\approx 20$ km s$^{-1}$, which
is close to the dispersion of the stellar velocities at the distance
of $R_0$. The dispersion of the initial  radial velocities in the
Hercules region is $\sigma_R\approx 30$ km s$^{-1}$, which is also
close to the dispersion of the radial velocities in the surrounding
disk. In the anti-Hercules region, the dispersion of the initial
radial velocities is noticeably larger, $\sigma_R\approx 38$ km
s$^{-1}$, which is possibly due to orbital libration.

\begin{figure*}
\centering  \resizebox{14 cm}{!}{\includegraphics{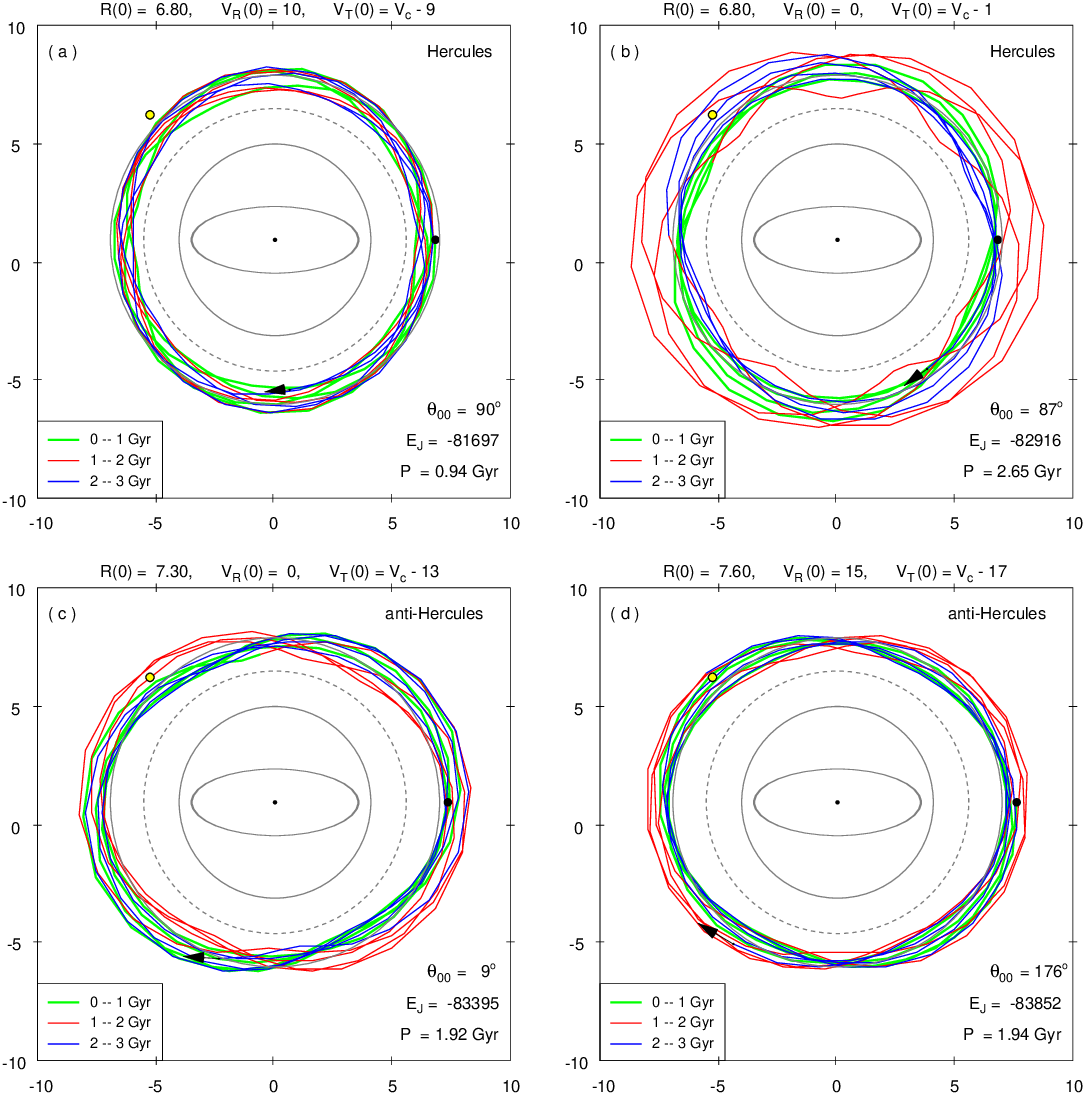}}
\caption{Typical orbits of stars in the Hercules (top row) and
anti-Hercules (bottom row) regions. The orbits are shown in the bar
reference frame, in which stars considered rotate in the direction
opposite that of the Galactic rotation. The supposed position of the
Sun and the initial position of the star are shown by the yellow and
black circles, respectively. The initial values of the distances,
$R(0)$, radial, $V_R(0)$, and azimuthal, $V_T(0)$, velocities of the
stars are given at the top of each frame.  The segments of the orbits
outlined by the stars at the time intervals  0--1, 1--2 and 2--3 Gyr
are shown in green, red and blue, respectively. For each orbit, we
list the angle $\theta_{00}$, which characterizes the average
orientation of the orbit relative to the major axis of the bar at the
time interval 0--3 Gyr, the Jacobi energy $E_J$, and the period $P$
of variations in the angular momentum. The values of distances,
velocities and $E_J$ are given in units of kpc, km s$^{-1}$ and
km$^2$ s$^{-2}$, respectively. Also are shown the positions of the
bar (ellipse), CR and OLR (solid gray lines), and $-4/1$ resonance
(gray dashed line). It is clearly seen that the orbits in the
Hercules region are, on average, elongated perpendicular to the bar
while the orbits in the anti-Hercules region are oriented parallel to
the bar.} \label{fig_orb_4}
\end{figure*}
\begin{figure*}
\centering  \resizebox{14 cm}{!}{\includegraphics{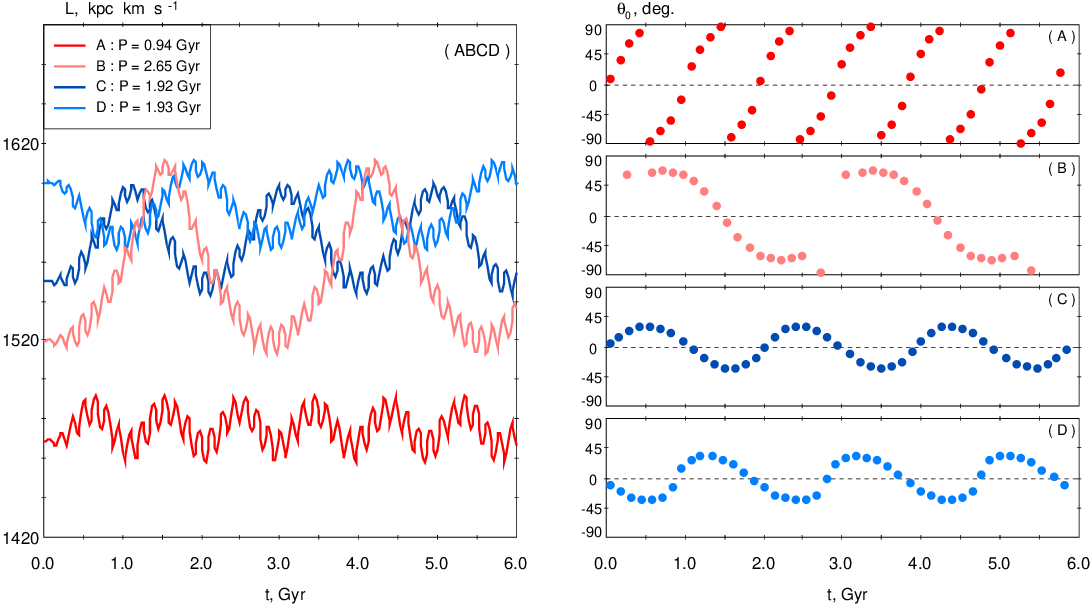}}
\caption{(Left panel) Oscillations of the angular momentum $L$ of the
stars A, B, C, and D, whose orbits are shown in  frames (a), (b),
(c), and (d) of Fig.~\ref{fig_orb_4}, respectively. We can clearly
see the fast and slow oscillations of $L$. The periods of the slow
oscillations of $L$ for stars $A$, $B$, $C$, and $D$ are  0.94, 2.65,
1.92, and 1.93 Gyr, respectively. (Right panel) Oscillations of the
angle $\theta_0$, which characterizes the direction of orbit
elongation at the time interval of one radial oscillation of the
star. It is clearly seen that in frame (A) the angle $\theta_0$ only
increases while it oscillates in frames (B), (C), and (D). }
\label{fig_L_theta}
\end{figure*}
\begin{figure*}
\centering  \resizebox{14 cm}{!}{\includegraphics{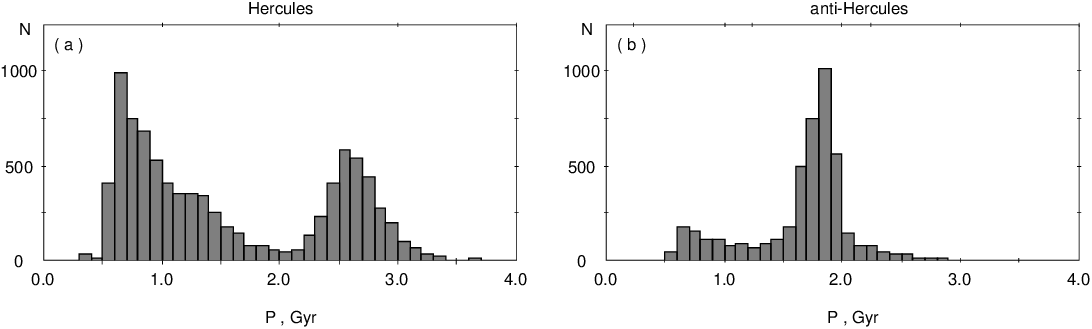}}
\caption{Distribution of orbits along the period $P$ of slow
variations in the angular momentum in the (a) Hercules and (b)
anti-Hercules regions. It is clearly seen that in the Hercules
region, the distribution has two maxima located at $P=0.7$ and 2.6
Gyr while in the anti-Hercules region, the maximum corresponds to
$P=1.9$ Gyr. } \label{fig_his_per}
\end{figure*}

\subsection{4.5 Orbits in the Hercules and anti-Hercules regions}

Fig.~\ref{fig_orb_4} shows typical stellar orbits in the Hercules
(upper row) and anti-Hercules (lower row) regions. The Galaxy rotates
counterclockwise. The orbits are considered in the reference frame of
the rotating bar. All  stars considered are located outside the
Corotation Radius of the bar ($R>R_{CR}$), so in the bar reference
frame they rotate in the sense opposite that of the Galactic
rotation, i.~e. clockwise. The supposed position of the Sun relative
to the bar and the initial position of the star are shown by the
yellow and black circles, respectively. The initial values of the
distances, $R(0)$, radial, $V_R(0)$, and azimuthal, $V_T(0)$,
velocities of the stars are given at the top of each frame.  The
segments of the orbits at the time intervals  0--1, 1--2 and 2--3 Gyr
are shown in green, red and blue, respectively.  For each orbit, we
give the angle $\theta_{00}$, which determines  the average
orientation of the orbit relative to the major axis of the bar at the
time interval 0--3 Gyr. We will also consider the angle $\theta_0$
characterizing the orientation of the orbit relative to the major
axis of the bar at the time interval of one radial oscillation of a
star (from one intersection of the average radius of the orbit with a
negative radial velocity to another). For each star,  we present the
period $P$ of variations in the angular momentum and total energy, as
well as the Jacobi energy $E_J$, which is conserved after the bar
reaches its full power.

Fig.~\ref{fig_L_theta} (left panel) shows oscillations of the angular
momentum $L$ of the stars $A$, $B$, $C$, and $D$, whose orbits are
shown in frames (a), (b), (c), and (d) of Fig.~\ref{fig_orb_4},
respectively. The fast and slow oscillations of $L$ are clearly
visible. The fast oscillations with a period of $P=0.13$ Gyr occur
twice during one period of the star's revolution relative to the bar.
The slow oscillations  result  from the beats between the frequency
with which a star encounters perturbations from the bar,
$2(\Omega-\Omega_b)$, and the epicyclic frequency, $\kappa$
\citep[for more details, see][]{melnik2023, melnik2024}. The periods
of  slow oscillations in $L$ of stars $A$, $B$, $C$, and $D$ are
$P=0.94$, 2.65, 1.92, and 1.93 Gyr, respectively.

Fig.~\ref{fig_L_theta} (right panel) shows oscillations of the angle
$\theta_0$ characterizing the direction of orbit elongation at the
time interval of one radial oscillation. Frames (A), (B), (C), and
(D) correspond to the stellar orbits shown in frames (a), (b), (c),
and (d) of Fig.~\ref{fig_orb_4}. It is clearly seen that the periods
$P$ of slow variations in the angular momentum $L$ coincide with the
periods of variations in the direction of the orbit elongation. Note
that the eccentricity and average radius of orbits also change with
the period $P$ \citep[see, for example,][Fig.~11]{melnik2023}.

Fig.~\ref{fig_orb_4}a shows a typical orbit of a star in the Hercules
region. The initial distance is $R(0)=6.8$ kpc, which is less than
the OLR radius ($R_{OLR}=7.0$ kpc), and the initial azimuthal
velocity is $V_T(0)=216$ km s$^{-1}$, which is slightly less than the
most probable value (218 km s$^{-1}$) of the initial azimuthal
velocity at this distance. The initial radial velocity is $V_R(0)=10$
km s$^{-1}$. A small positive radial velocity is needed for the orbit
to be elongated enough to bring the star into the solar vicinity of
$r<0.5$ kpc. It is clearly seen that the orbit changes with time, but
it always lies inside the figure bounded by two ellipses elongated
perpendicular to the bar. The average value of the angle
$\theta_{00}$ is $\theta_{00}=90^\circ$.

Fig.~\ref{fig_L_theta}A shows the change in the direction of
elongation of the star's orbit (Fig.~\ref{fig_orb_4}a) over time. We
can  clearly see that the angle $\theta_0$ increases almost linearly
from the value $-90$ to $90^\circ$, and then a new revolution begins.
The values of $\theta_0=\pm90^\circ$ correspond to the same
orientation of the elliptical orbit. The change in the direction of
the orbit elongation of the star $A$ occurs only in the direction of
increasing $\theta_0$. Thus, here we  are not dealing with
oscillations of orbit orientation but with the rotation of the
direction of  orbit elongation in one sense. In the case considered,
the orbit rotates in the positive sense, i.~e. in the sense  of the
Galactic rotation. Note that the rotation velocity of the orbit
increases slightly when the orbit is oriented parallel to the bar
($\theta_0\approx 0^\circ$).

Fig.~\ref{fig_orb_4}b shows another typical orbit in the Hercules
region. The initial values of the distance and azimuthal velocity of
the star considered are $R(0)=6.8$ kpc and $V_T(0)=224$ km s$^{-1}$.
It is clearly seen that the direction of  orbit elongation changes
from $\theta_0=\pm90^\circ$ (elongated perpendicular to the bar) to
$\theta_0=0^\circ$ (elongated parallel to the bar). The angle
$\theta_{00}$, which determines the average orientation of the orbit,
is $\theta_0=87^\circ$, because  most of the time at the interval
0--3 Gyr the orbit is elongated perpendicular to the bar.
Fig.~\ref{fig_L_theta}B shows that oscillations of the angle
$\theta_0$ are not sinusoidal. We can see a sharp jump from the value
$\theta_0=-90^\circ$ to $60^\circ$, then $\theta_0$ decreases
sinusoidally to $-60^\circ$, after which there is a rapid
rearrangement of the orbit: first the value $\theta_0$ drops to
$-90^\circ$,  then sharply increases to $\theta_0=60^\circ$. Note
that the Galactocentric distance $R$ to the star varies in a very
wide range, $R=6.0$--8.8 kpc, i.e. the star comes very close to the
resonance $-4/1$ ($R_{-4/1}=5.5$ kpc) in the inner region and goes to
distances exceeding the OLR radius ($R_{OLR}=7.0$ kpc) by 1.8 kpc in
the outer region. In general, we view  an orbit with a wide range of
the distance $R$ variation.

Possibly,  it is these orbits (Fig.~\ref{fig_orb_4}b) that increase
the number of stars $N_H$ in the Hercules region at the period  2--3
Gyr (Fig.~\ref{fig_time_n}a). It is clearly seen that the angle
$\theta_0$ at the time interval of 1.8--2.6 Gyr  has a value close to
$\theta_0 \approx -60^\circ$ (Fig.~\ref{fig_L_theta}B). In this case,
the orbit is tilted to the left (in the sense of the Galactic
rotation) relative to the minor axis of the bar and lies near the Sun
(Fig.~\ref{fig_orb_4}b, blue line), besides, the star passes the Sun
with a positive radial velocity, and, consequently, can fall inside
the Hercules region on the  ($V_R$, $V_T$) plane. In other time
intervals, this star does not fall into the solar neighborhood,
$r<0.5$ kpc, with a positive radial velocity ($10<V_R<40$ km
s$^{-1}$): at the time period  0.5--1.3 Gyr, its orbit is inclined to
the right relative to the minor axis of the bar
(Fig.~\ref{fig_orb_4}b, green line) and lies  at a large distance
from the Sun while at the time period  1.3--1.8 Gyr, its orbit is
stretched almost parallel to the bar ($\theta_0 \approx 0^\circ$) and
the star flies near the Sun with a negative radial velocity
(Fig.~\ref{fig_orb_4}b, red line).

Fig.~\ref{fig_orb_4}c shows a typical orbit of a star in the
anti-Hercules region. The initial values of the distance and
azimuthal velocity are $R(0)=7.3$ kpc and $V_T(0)=212$ km s$^{-1}$.
It is clearly seen that the orbit is elongated, on  average, almost
parallel to the bar, the angle $\theta_{00}$ is
$\theta_{00}=9^\circ$. The fact that the value $\theta_{00}$ is not
exactly zero, $\theta_{00}\neq 0^\circ$, is due to the circumstance
that most of the time during the interval 0--3 Gyr, this orbit is
tilted to the right relative to the minor axis of the bar. The
direction of orbit elongation oscillates almost sinusoidally from
$-45$ to 45$^\circ$ with the period $P=1.92$ Gyr
(Fig.~\ref{fig_L_theta}C). In our previous works, we showed that
oscillations of such orbits cause the periodic change in the
morphology of the outer rings and the appearance of humps on the
profiles of the $V_R$-velocity distribution  along the distance $R$
\citep{melnik2023,melnik2024}.

Fig.~\ref{fig_orb_4}d shows another orbit of a star in the
anti-Hercules region. The initial values of the distance and
azimuthal velocity are $R(0)=8.0$ kpc and $V_T(0)=200$ km s$^{-1}$.
The angle $\theta_{00}$ is $\theta_{00}=176^\circ$, i.~e.  the orbit
is tilted to the left relative to the minor axis of the bar most of
the time at the interval 0--3 Gyr. It is clearly seen that the angle
$\theta_0$ slowly decreases from 45$^\circ$ to $-45^\circ$ and then
quickly grows back to 45$^\circ$ (Fig.~\ref{fig_L_theta}D).

A comparison of Fig.~\ref{fig_L_theta}C and Fig.~\ref{fig_L_theta}D,
as well as Fig.~\ref{fig_orb_4}c and Fig.~\ref{fig_orb_4}d, shows
that oscillations in the direction of orbit elongation of  the stars
$C$ and $D$ occur practically in antiphase.

In general, the orbits of stars in the Hercules region
(Fig.~\ref{fig_orb_4}a and Fig.~\ref{fig_orb_4}b) support the outer
ring $R_1$ elongated perpendicular to the bar, while in the
anti-Hercules region (Fig.~\ref{fig_orb_4}c and
Fig.~\ref{fig_orb_4}d) -- the ring $R_2$ elongated parallel to the
bar.

The eccentricities of the orbits shown in Fig.~\ref{fig_orb_4} also
vary with the periods equal to the periods of slow variations in the
angular momentum of  the stars. The ranges of variations in the
eccentricities of the stars $A$, $B$, $C$, and $D$ are (a)
0.38--0.58, (b) 0.44--0.73, (c) 0.43--0.70, and (d) 0.30--0.62,
respectively. Thus, the stars $B$ and $C$ have  more elongated orbits
than stars $A$ and $D$.

\subsection{4.6 Distribution of orbits along the  period $P$ in the Hercules and anti-Hercules regions}

Fig.~\ref{fig_his_per} shows the distribution of orbits along the
period $P$ of  slow variations in the angular momentum in the (a)
Hercules and (b) anti-Hercules regions. It is clearly seen that in
the Hercules region, the distribution has two maxima corresponding to
the periods $P=0.7$ and 2.6 Gyr. The first maximum ($P=0.7$ Gyr) is
created by orbits that always lie inside the figure bounded by two
ellipses stretched perpendicular to the bar (Fig.~\ref{fig_orb_4}a),
therefore, these orbits are, on average, stretched perpendicular to
the bar, and stars on them spend most of their time inside the OLR
radius, $R \le R_{OLR}$. The second maximum ($P=2.6$ Gyr) is created
by orbits that are most of the time elongated at the angles $\theta_0
\approx 60^\circ$ or $\theta_0 \approx -60^\circ$ to the major axis
of the bar, and a small part of their time -- parallel to the bar
(Fig.~\ref{fig_orb_4}b). In addition, these orbits are characterized
by a large range of variation in the distance $R$, $\Delta R>2.5$
kpc. In this case, stars are located both inside and outside the OLR
radius approximately half of the time.

Fig.~\ref{fig_his_per}b shows that the  maximum of the orbit
distribution along the period $P$ in the anti-Hercules region
corresponds to  $P=1.9$ Gyr. This period coincides with the most
probable value of the period of the angular momentum variations of
stars on librating orbits near the OLR \citep{melnik2024}. Thus, most
orbits in the anti-Hercules region are expected to librate relative
to the major axis of the bar.

\begin{table*}
 \caption{Median $G$, $G-G_{RP}$ and $M_{G}$ in the
Hercules and anti-Hercules regions from the Gaia DR3 data }
 \centering
\begin{tabular}{ccc|cc}
\hline
\\
& \multicolumn{2}{c|}{Hercules} & \multicolumn{2}{c}{anti-Hercules}\\
\\
\multicolumn{5}{c}{\hspace{0.1cm} $G$}\\
\\
         &   $13.4174  \pm0.0042^m$ & 148403 & $13.4972 \pm0.0048^m$ & 108002    \\
\\
\hline
\\
\multicolumn{5}{c}{\hspace{0.3cm} $G-G_{RP}$}\\
 \\
         &   $0.6682  \pm0.0005^m$ & 147903 &   $0.6792 \pm0.0006^m$ & 107656  \\
\\
\hline
\\
\multicolumn{5}{c}{\hspace{0.2cm} $M_{G}$}\\
\\
          &   $5.7723  \pm0.0045^m$ & 148403 &   $5.9110 \pm0.0054^m$ & 108002   \\
\\
\hline
\end{tabular}
\label{tab:statistics}
\end{table*}

\subsection{4.7 Comparison with the Gaia DR3 data}

\begin{figure*}
\centering  \resizebox{14 cm}{!}{\includegraphics{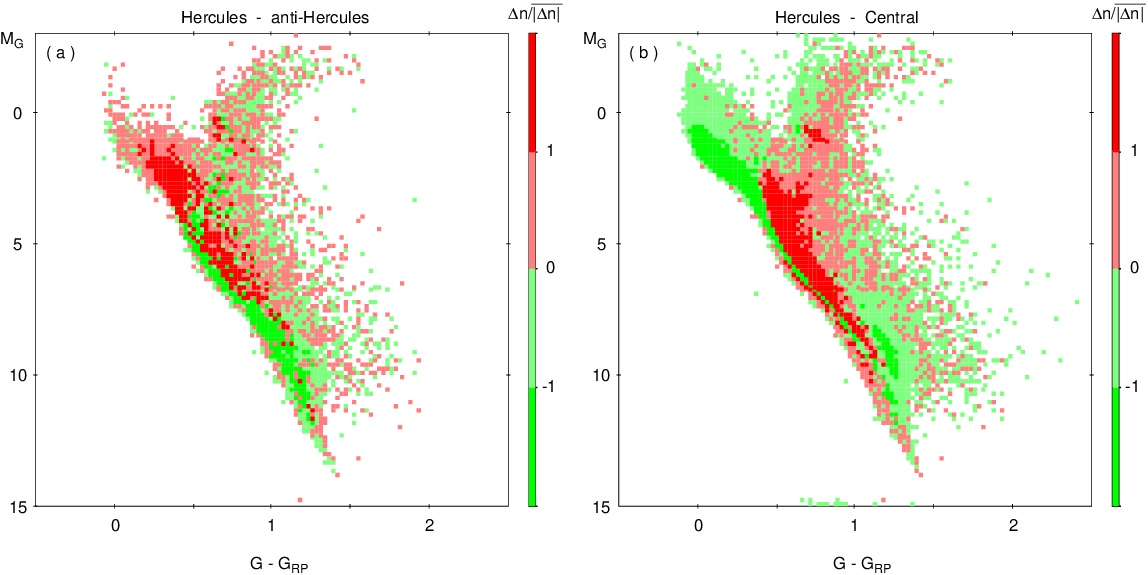}}
\caption{Distribution of the differences $\Delta n$ between the
number of Gaia DR3 stars in the Hercules region and the normalized
number of stars in (a) the anti-Hercules region and (b) the Central
region on the Hertzsprung-Russell diagram. The horizontal axis shows
the color, $G-G_{RP}$, and the vertical axis indicates the absolute
stellar magnitude, $M_G$. The sizes of the areas are 0.026 and
0.157$^m$, respectively. Areas in which the number of stars in the
Hercules region is greater (less) than the normalized number of stars
in another region (namely, (a) the anti-Hercules region and (b) the
Central region) are shown in shades of red (green). The average value
of the difference $\Delta n$ is $\overline{|\Delta n|}\approx 7$. The
excess of stars in one region over the number of stars in another
region by more than $\overline{|\Delta n|}$ is shown in bright red or
green, respectively (see the color palette on the right). (a) It is
clearly seen that the upper part of the main sequence corresponding
to the bluest and most luminous stars is colored red indicating that
the stars of the Hercules region predominate over those of the
anti-Hercules region in this part of the diagram while the red giant
branch is represented by both regions nearly equally. (b) The Central
region includes relatively bluer and more luminous main-sequence
stars than the Hercules region, and the red giant branch is
represented mainly by stars of the Hercules region.}
\label{dif_H_aH_c}
\end{figure*}

\begin{figure*}
\centering  \resizebox{13 cm}{!}{\includegraphics{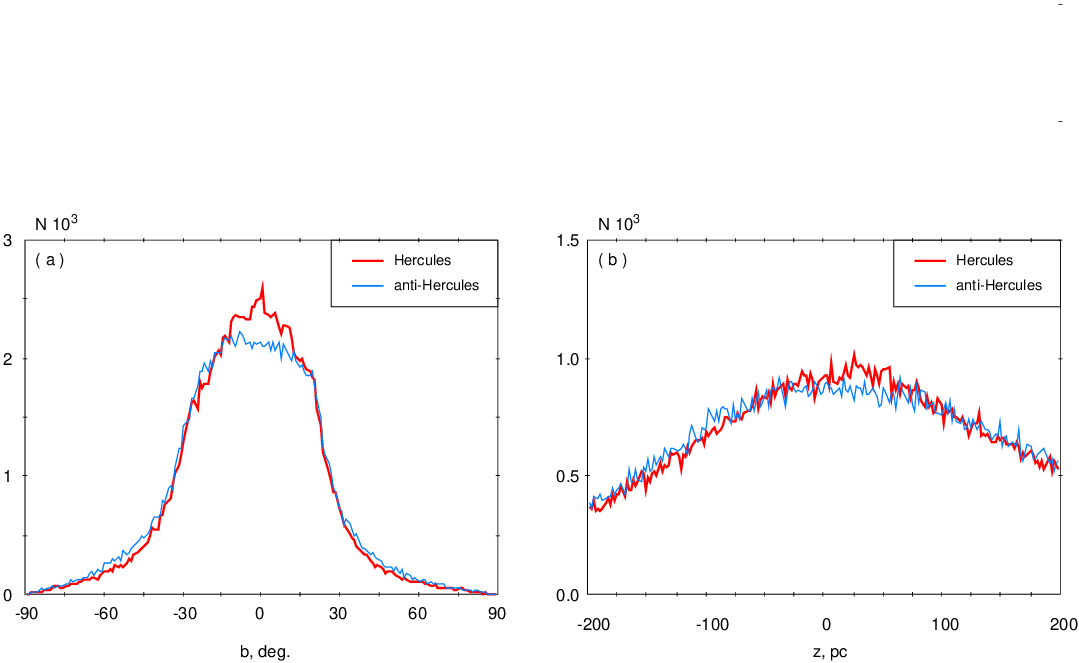}}
\caption{Normalized distributions of the number of stars $N$ in the
Hercules (red line) and anti-Hercules (blue line) regions along (a)
the Galactic latitude $b$ and (b) the Galactic coordinate $z$. The
width of the bins is (a) $\Delta b=1^\circ$ and (b) $\Delta z=2$ pc.
(a) It is clearly seen that the Hercules region includes a
significant excess of stars near the Galactic plane compared to the
anti-Hercules region, which is, on average, 11\% in the range
$|b|<10^\circ$ and reaches maximum value of 22\% at $b=0$. The
distributions of stars in both regions are asymmetric with respect to
the $b=0$ line,  and $N$ drops much faster toward positive latitudes
$b$. (b) We can also see the excess of stars of the Hercules region
near the Galactic plane compared to the anti-Hercules one, which is,
on average, 10\% at the interval $z=14$--54 pc. Maximum value of $N$
in the Hercules region corresponds to $z=24$ pc. The distribution of
stars in the anti-Hercules region has a broad maximum corresponding
to the interval $z \in [-46, 66]$ pc.} \label{n_b}
\end{figure*}

\begin{figure*}
\centering  \resizebox{12 cm}{!}{\includegraphics{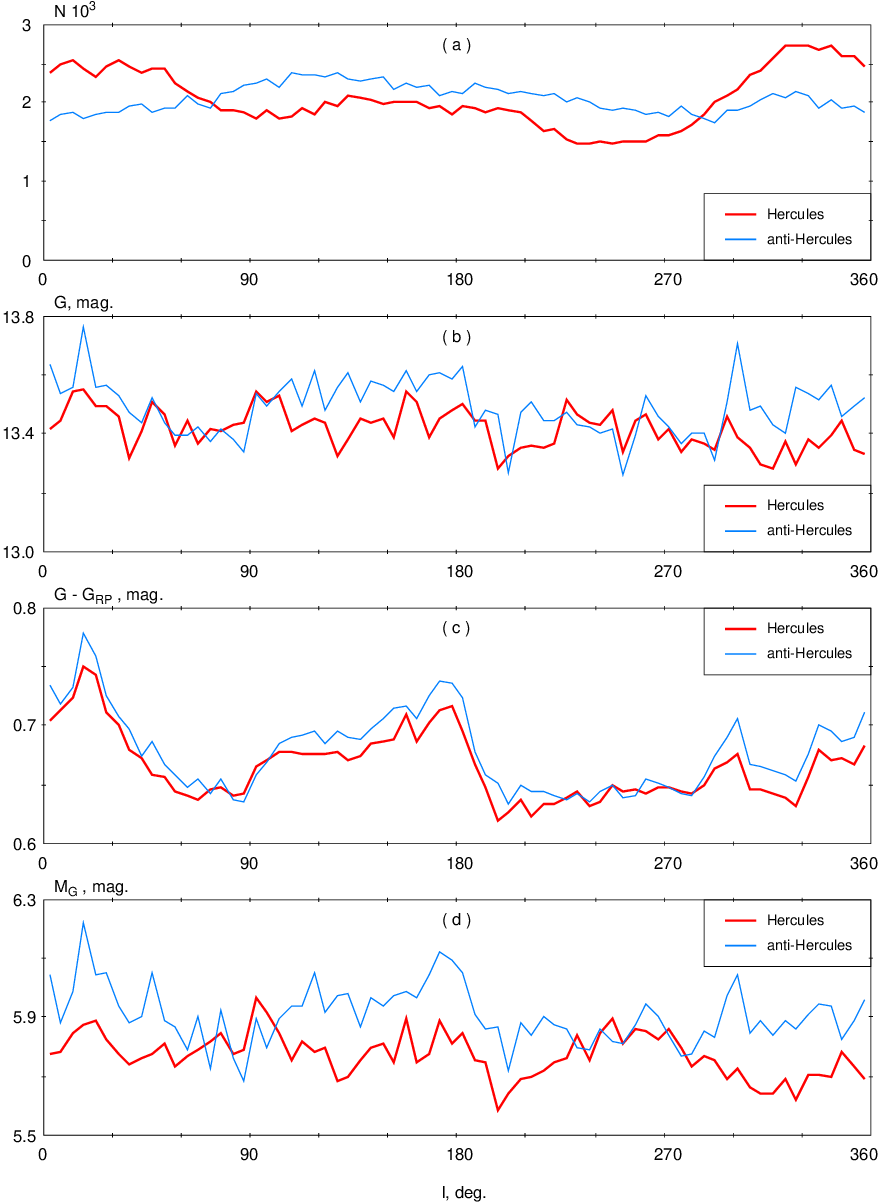}}
\caption{Distributions of the (a) number of stars, $N$, (b)
magnitude, $G$, (c)  color, $G-G_{RP}$, and (d)  absolute magnitude,
$M_G$, of  stars in the Hercules (red line) and anti-Hercules (blue
line) regions along the Galactic longitude $l$, calculated from Gaia
DR3 data. The values of $N$ and the median values of $G$, $G-G_{RP}$,
and $M_G$ were calculated in sectors of the $\Delta l=5^\circ$ width.
(a) It is clearly seen that the Hercules region contains relatively
more stars in the direction toward the Galactic center, $l<60\cup
l>300^\circ$, than the anti-Hercules region. (b) In the direction
toward the Galactic center, $l<45 \cup l>300^\circ$, and in the
second quadrant, $90 < l<180^\circ$, stars of the Hercules region
are, on average, brighter than those of the anti-Hercules region
while  in other directions the brightness of stars in both regions is
approximately the same. (c) Almost everywhere, stars of the Hercules
region are, on average, bluer in color than those of the
anti-Hercules region. (d) The Hercules region includes more luminous
stars than the anti-Hercules one almost everywhere.} \label{n_l}
\end{figure*}


The study of model-star  samples  has shown that at the initial time,
stars in the Hercules region are, on average, slightly closer to the
Galactic center than stars in the anti-Hercules region (Section 4.4).
If the Galactic disk formed inside-out
\citep[e.~g.,][]{chiappini2001} then we can expect that stars in the
Hercules region must be, on average, older and redder than stars in
the anti-Hercules region. Here we use the fact that most of stars in
the Gaia catalog lie on the main sequence of the Hertzsprung-Russell
diagram \citep{babusiaux2018}. This hypothesis would be correct if
the overwhelming majority of stars in the solar neighborhood were
formed before the epoch of the bar formation, but recent episodes of
star formation can alter the distribution of stellar ages in these
regions. In addition,  selection effects depend not only on the value
of the magnitude $G$ and color, but also on the direction to a star.
In crowded areas of the sky, the processing program automatically
selects brighter objects \citep[e.g.,][]{boubert2020}.

Table 1 presents the median values of the  magnitude, $G$, color,
$G-G_{RP}$, and absolute magnitude, $M_{G}$, of stars in the Hercules
and anti-Hercules regions derived from the Gaia DR3 data. It also
provides their uncertainties  and the number of stars with known
values of the corresponding parameters. The absolute magnitudes $M_G$
were calculated without  taking into account the extinction, which
must be small in the vicinity of 0.5 kpc from the Sun:

\begin{equation}
M_G= G-5\log_{10} r -10, \label{M_G}
\end{equation}

\noindent where $r$ is the heliocentric distance in units of kpc. We
use the $G-G_{RP}$ values because they are more accurate than
$G_{BP}-G_{RP}$ and  are measured for a larger number of stars
\citep[][]{andrae2018,vallenari2023}.

Table 1 shows that the Hercules region includes slightly brighter,
bluer, and more luminous  stars than the anti-Hercules region.
Although the differences in the median values do not exceed $0.14^m$,
their statistical significance (the ratio of the value to its
uncertainty) exceeds $8\sigma$.

Fig.~\ref{dif_H_aH_c}a shows the distribution of the differences
$\Delta n$ between  the number of Gaia DR3 stars in the Hercules
region and the normalized number of stars in the anti-Hercules
region, calculated in small areas on the Hertzsprung-Russell diagram.
The horizontal axis shows the color, $G-G_{RP}$, while the vertical
axis indicates the absolute stellar magnitude $M_{G}$. The sizes of
the areas are 0.026 and 0.157$^m$, respectively. Normalization to the
number of stars is necessary, as the Hercules and anti-Hercules
regions include the different numbers of stars. The normalization
factor, $k=1.37$, equals the ratio of the numbers of stars with known
line-of-sight velocities, parallaxes, $G$ and $G_{RP}$ values in the
Hercules and anti-Hercules regions. Areas in which the number of
stars in the Hercules region exceeds (falls below)  the normalized
number of stars in the anti-Hercules region are shown in shades of
red (green), respectively. The average value of the difference
$\Delta n$ is $\overline{|\Delta n|} \approx 7$. It is clearly seen
that the part of the main sequence corresponding to the bluest and
most luminous stars is colored red, indicating an excess of stars in
the Hercules region compared to the anti-Hercules region in this part
of the diagram. Note that the red giant branch is represented almost
equally by both regions. Therefore, the Hercules region likely
contains a larger fraction of relatively young stars than the
anti-Hercules region. However, this result may also be due to
selection effects: stars in the Hercules and anti-Hercules regions
are distributed differently in space, although they are enclosed by
the surfaces $r<0.5$ and $|z|<0.2$ kpc, which will be discussed
below.

For completeness, we also compared the distributions of stars in the
Hercules and Central regions (Fig.~\ref{dif_H_aH_c}b). The Central
region includes 361617 Gaia DR3 stars located in the solar
neighborhood, $r<0.5$ and $|z|<0.2$ kpc,  that satisfy the criterion
$|V_z|<50$ km s$^{-1}$, and lie on the  ($V_R$, $V_T$) plane inside
the ellipse  centered at $V_R=0$ and $V_T=225$ km s$^{-1}$ with
semi-axes $\Delta V_R=15$ and $\Delta V_T=12$ km s$^{-1}$. Since the
velocity of the rotation curve at the solar distance is $V_c=225$,
the Central region contains a larger portion of stars with velocities
close to that of the rotation curve, therefore, this region likely
contains a higher proportion of young stars. In this case, the
normalization factor is $k=0.41$. It is clearly seen that the upper
part of the main sequence is colored green, i.e. the Central region
indeed contains relatively more blue stars than the Hercules region.
On the other hand, the red giant branch is represented mainly by
stars of the Hercules region. Consequently, the fraction of young
stars in the Central region is indeed larger than in the Hercules
region.

However, the differences in the distribution of stars in the Hercules
and anti-Hercules regions on the Hertzsprung-Russell diagram may
still be caused by selection effects. The selection function of Gaia
DR3 stars into different samples, for example, into a sample of stars
with known line-of-sight velocity, has a low value in the Galactic
plane, and especially in the direction of the Galactic center. This
indicates a lack of stars ($G<21^m$) with measured line-of-sight
velocity in these directions \citep{castro-ginard2023}. Since the
Gaia DR3 stars in the Hercules and anti-Hercules regions are
distributed differently in the sky, the influence of selection
effects on these samples may be different.

Fig.~\ref{n_b}a shows the distributions of the numbers of stars $N$
in the Hercules and anti-Hercules regions as a function of the
Galactic latitude $b$. Both distributions were normalized to the
number of stars in the Hercules region, $N_H=148404$, i.e. the number
of stars in each sector in the anti-Hercules region was multiplied by
the factor $k=1.37$. The width of the sectors is $\Delta b=1^\circ$.
It is clearly seen that the Hercules region contains a significant
excess of stars near the Galactic plane compared to the anti-Hercules
region, which, on average, equals 11\% in the range $|b|<10^\circ$
and reaches maximum value of 22\% at $b=0$. Probably, the weaker
concentration of stars of the anti-Hercules region toward the
Galactic plane allowed the Gaia spectrometer to measure line-of-sight
velocities for a larger fraction of  faint stars in this region
compared to the Hercules region. Thus, the number of stars in the
anti-Hercules region $N_{aH}$ was increased at the expense of faint
stars, which decreased the relative abundance of bright stars in this
sample. Note also that the distributions of stars in both regions are
asymmetric with respect to the $b=0$ line, and $N$ decreases much
faster in the direction of positive $b$ values than in the opposite
direction.

Fig.~\ref{n_b}b shows the distributions of the numbers of stars $N$
in the Hercules and anti-Hercules regions normalized to the number of
stars in the Hercules region along the Galactic coordinate $z$. The
bin width is $\Delta z=2$ pc. To calculate $z$, we used the value of
the solar coordinate $z_\odot$ equal to $z_\odot=27$ pc
\citep{bland2016}. Here we also see an excess of stars of the
Hercules region near the Galactic plane compared to the anti-Hercules
region, which,  on average, is 10\% at the interval $z=14$--54 pc.
Maximum value of the number of stars $N$ in the Hercules region
corresponds to $z=24$ pc. The distribution of stars in the
anti-Hercules region has a broad maximum corresponding to the
interval $z \in [-46, 66]$ pc.

Fig.~\ref{n_l} shows the distributions of (a) the number of stars,
$N$, (b) magnitude, $G$, (c) color, $G-G_{RP}$, and (d) absolute
magnitude, $M_G$, of Gaia DR3 stars located in the Hercules and
anti-Hercules regions, along the Galactic longitude $l$. For each
longitude sector of the $\Delta l=5^\circ$ width, we calculated the
number of stars $N$ within the sector, as well as the median values
of $G$, $G-G_{RP}$ and $M_G$.

Fig.~\ref{n_l}a shows the distribution of the number of stars $N$
along the longitude $l$ normalized by the number of stars in the
Hercules region. We can clearly see that the Hercules region contains
relatively more stars in the direction toward the Galactic center, $l
< 60 \cup l>300^\circ$, than the anti-Hercules region while
anti-Hercules region has an excess of stars in other directions,
$60<l<300^\circ$. The excess of stars of the Hercules region in the
direction toward the Galactic center may be related to the orbits
shown in Fig.~\ref{fig_orb_4}a which cross the solar neighborhood,
$r<0.5$ kpc, at the edge closest to the Galactic center. We carried
out experiments by excluding stars lying in different sectors of the
longitude $l$, including the central sector, $l<60\cup l>300^\circ$,
but this did not result in the disappearance of the excess of stars
of the Hercules region at the top of the main sequence
(Fig.~\ref{dif_H_aH_c}a).

Fig.~\ref{n_l}b shows the distributions of the median  magnitude $G$
of stars in the Hercules and anti-Hercules regions calculated in
sectors along the longitude $l$. It is clearly seen that in the
direction toward the Galactic center, $l<45 \cup l>315^\circ$, and in
the  second quadrant, $90 < l<180^\circ$,  stars in the Hercules
region are, on average, brighter than in the anti-Hercules region
while in other directions, the brightness of the stars in both
samples is approximately the same.

Fig.~\ref{n_l}c shows the  distributions of the color of stars,
$G-G_{RP}$, in the Hercules and anti-Hercules regions. Interestingly,
stars in both regions are redder in the direction toward the Galactic
center, $l<45 \cup l>315^\circ$, and in the second quadrant, $90 <
l<180^\circ$, while in the  third quadrant, $180 < l<270^\circ$,
there is a sharp blueing of stars of both samples. Note, that almost
everywhere, stars in the Hercules region are, on average, bluer than
in the anti-Hercules region.

Fig.~\ref{n_l}d shows the distributions of the absolute  magnitude,
$M_G$. It is clearly seen that almost everywhere the Hercules region
includes more luminous stars than the anti-Hercules region. The
exceptions are two sectors: $70 < l<110^\circ$ and $230 <
l<290^\circ$, where the median  $M_G$ values are nearly the same in
both regions. Note also that  stars in the Hercules region are
located, on average, farther away from the Sun than in the
anti-Hercules region, the median estimates of the heliocentric
distances $r$ in these regions are $r=338.7\pm0.3$ and $321.5\pm0.4$
pc, respectively.

The differences in the brightness, color, and luminosity of stars in
the Hercules and anti-Hercules regions are possibly caused by
different concentrations of these stars in the sky relative to the
Galactic plane, which leads to an underestimation of the number of
faint stars in the Hercules region compared to the anti-Hercules
region. In general, the lack of stars with measured line-of-sight
velocities in the Hercules region compared to the anti-Hercules one
must result in  an underestimation of the $f_g$ value derived from
the Gaia DR3 data (Eq.~\ref{frac}).

\section{5. Conclusions}

We  studied the formation of the Hercules stream using the model of
the Galaxy with an analytical bar. The model disk forms an outer
resonance ring $R_1R_2$ located near the OLR of the bar. This model
reproduces well the distributions of the radial, $V_R$, and
azimuthal, $V_T$, stellar velocities along the Galactocentric
distance, $R$, derived from the Gaia EDR3 and Gaia DR3 data. The
positions of the Sun and the OLR radius of the bar correspond to the
distances of $R_0=7.5$ and $R_{OLR}=7.0$ kpc
\citep{melnik2021,melnik2023}.

We analyzed  variations in the number of model stars in the Hercules
and anti-Hercules regions  over time (Fig.~\ref{fig_vr_vt_4}). We
need the anti-Hercules region for the calibration of the number of
stars. A strict definition of the Hercules and anti-Hercules regions
includes two criteria. First, during the time period considered,
stars must fall into the solar neighborhood, $r<0.5$ kpc. Second,
these stars must have  velocities, $V_R$ and $V_T$, that lie on the
($V_R$, $V_T$) plane inside ellipses centered at $V_R=25$ and
$V_T=200$ km s$^{-1}$ (Hercules), or at $V_R=-25$ and $V_T=200$
(anti-Hercules), with semi-axes of $\Delta V_R=15$ and $\Delta
V_T=12$ km s$^{-1}$. The velocities $V_R$ and $V_T$ are calculated
relative to the Galactic center.

The number of stars in the Hercules region, $N_H$, reaches its first
maximum at the time period of 2.0--2.6 Gyr from the start of
simulation and then decreases. The second maximum $N_H$ corresponds
to the period of 4.3--5.2 Gyr and looks more like a plateau. It is
interesting that the number of stars $N_H$ reaches its maximum
precisely at those time periods when the humps on the $V_R$-velocity
distributions  along $R$ disappear ($2.5\pm0.3$ and $4.5\pm0.5$ Gyr),
and the model best agrees with observations \citep{melnik2024}.

The number of stars in the anti-Hercules region, $N_{aH}$,
demonstrates well-defined oscillations with the period of
$P=1.8\pm0.1$ Gyr. These oscillations are probably related to orbital
librations near the OLR with the period close to $P\approx 2.0$ Gyr
\citep{melnik2023, melnik2024}.

We calculated the relative difference, $f$, in the number of stars in
the Hercules and anti-Hercules regions (Eq.~\ref{frac}), which does
not depend on the number of stars in the model and observational
samples. The value of $f$ calculated for model-disk stars oscillates
in the range $[-0.009, 0.656]$, and its average value is
$\overline{f}=0.291\pm0.011$, which is quite close to the value of
$f_g=0.315\pm0.004$ obtained for stars from the Gaia DR3 catalog
(Fig.~\ref{fig_time_n}). Thus, there are many times when the model
values of $f$ coincide with the observational value  $f_g$. On the
other hand, contamination of the observational sample with thick-disk
and halo stars may lead to an underestimation of the $f_g$ value.

We investigated the orientation of  orbits of  model-disk stars in
the Hercules and anti-Hercules regions at the time period
$t=2.5$--2.6 Gyr from the start of  simulation. We divided  model
stars into three groups depending on the orientation of their orbits:
elliptical orbits, i.~e. orbits with the order of symmetry  $n_s=2$,
elongated perpendicular to the bar ($75\le\theta_{00}<105^\circ$);
elliptical orbits ($n_s=2$) elongated parallel to the bar
($0\le\theta_{00}<15^\circ$ or $165\le\theta_{00}< 180^\circ$); and
other orbits. The angle $\theta_{00}$ determines the average
orientation of the orbit relative to the major axis of the bar during
the time period $t=0$--3 Gyr. It turned out that in the Hercules
region 84\% of orbits have an  elliptical shape and are elongated
perpendicular to the bar while in the anti-Hercules region 68\% of
orbits have $n_s=2$ and are elongated parallel to the bar
(Fig.~\ref{fig_orient}, ~\ref{fig_his_theta}). This strongly differs
from the distribution of all stars that fall in the solar vicinity,
$r<0.5$ kpc, during the period considered: 21\% have elliptical
orbits elongated perpendicular to the bar, and 19\% have elliptical
orbits elongated parallel to the bar.

We studied the distribution of stars in the Hercules and
anti-Hercules regions along the initial distances $R(0)$, as well as
variations  in the initial radial, $V_R(0)$, and azimuthal, $V_T(0)$,
velocities with variations in $R(0)$ (Fig.~\ref{fig_his_r0}). The
median values of the initial distances $R(0)$ in the Hercules and
anti-Hercules regions are 6.8 and 7.3 kpc, respectively, i.e. most
stars arrive in the Hercules region from the distances smaller than
the OLR radius  ($R_{OLR}=7.0$ kpc) and in the anti-Hercules region
from distances larger than $R_{OLR}$. The median values of the
initial azimuthal velocity $V_T$ calculated in  bins along $R(0)$
decrease with increasing $R(0)$, although there is a small plateau
(7.0--8.0 kpc)  in the Hercules region. As for the initial radial
velocity, the bins with  maximum number of particles $N$ correspond
to the velocity  $V_R(0)\approx 0$, which is in good agreement with
the fact that the most probable value of the initial radial velocity
in the model disk is zero.

In our previous papers, we  showed that the period $P$ of slow
variations in the  angular momentum, $L$, and total energy, $E$, of a
star coincides with the period of variations in the direction of
orbit elongation, mean size and eccentricity of the orbit
\citep{melnik2023,melnik2024}.

In the Hercules region, there are two types of orbits
(Fig.~\ref{fig_orb_4}a and Fig.~\ref{fig_orb_4}b). Orbits of the
first type (Fig.~\ref{fig_orb_4}a) change orientation, size, and
eccentricity with a period of $P \approx 0.7$ Gyr but always lie
inside a figure bounded by two ellipses elongated perpendicular to
the bar. The direction of elongation of these orbits shifts only in
the direction of the Galactic rotation (Fig.~\ref{fig_L_theta}A).
Orbits of the second type (Fig.~\ref{fig_orb_4}b) are elongated at
the angles $\theta_0 \approx -60^\circ$ or $60^\circ$ to the major
axis of the bar most of the time, and  parallel to the bar for a
small part of the time. These orbits librate relative to the major
axis of the bar in the range of angles $[-60, 60^\circ]$ with a
period of $P \approx 2.6$ Gyr (Fig.~\ref{fig_L_theta}B). Perhaps
these orbits cause an increase in the number of  stars, $N_H$, in the
Hercules region at the time period of 2--3 Gyr
(Fig.~\ref{fig_time_n}a).

The orbits in the anti-Hercules region librate relative to the major
axis of the bar in the range of angles $[-45, 45^\circ]$ with the
period of $P \approx 1.9$ Gyr (Fig.~\ref{fig_orb_4}c and
Fig.~\ref{fig_orb_4}d). Note that librations of the orbits shown in
Fig.~\ref{fig_orb_4}c cause periodic changes in the morphology of the
outer rings and the appearance of humps on the profiles of the
$V_R$-velocity distribution  along the distance $R$
\citep{melnik2023, melnik2024}.

On average, orbits in the Hercules region are elongated perpendicular
to the bar and support the outer ring $R_1$ while orbits in the
anti-Hercules region are elongated parallel to the bar and support
the outer ring $R_2$.

We investigated the distribution of the number of orbits along the
period $P$ of slow oscillations in the  angular momentum in the
Hercules and anti-Hercules regions (Fig.~\ref{fig_his_per}). In the
Hercules region, the distribution appears to have two maxima,
corresponding to $P=0.7$ and 2.6 Gyr. Probably, such a two-humped
distribution is connected with two types of orbits in the Hercules
region (Fig.~\ref{fig_orb_4}a and Fig.~\ref{fig_orb_4}b). In the
anti-Hercules region, the distribution has a well-defined maximum at
the period of $P=1.9$ Gyr, which coincides with the most probable
value of the period $P$ of librating orbits near the OLR
\citep{melnik2024}.

Analysis of the Gaia DR3 data showed that the Hercules region
includes an excess of stars in the upper part of the main sequence of
the Hertzsprung-Russell diagram compared to the anti-Hercules region
(Fig.~\ref{dif_H_aH_c}a, Table 1). This result may be caused by a
lack of faint stars in the Hercules region, which leads to an
increasing fraction of bright stars in this sample. The difference in
the relative abundance of  bright stars in the Hercules and
anti-Hercules regions may be due to the fact that  stars in the
Hercules region are more concentrated in the sky toward the Galactic
plane, which makes it difficult for the Gaia spectrometer to measure
their line-of-sight velocities (Fig.~\ref{n_b}). In general, the
selection effects  must also reduce the value of $f_g$ derived from
the Gaia DR3 data (Section 4.7).

\section*{Acknowledgements}

{\footnotesize We thank A.~K. Dambis for  fruitful discussion and
helpful suggestions. We thank the anonymous reviewer for helpful
comments and interesting discussion. The study was conducted under
the state assignment of Lomonosov Moscow State University. This work
was carried out using data from the European Space Agency (ESA)
mission {\it Gaia} (https:// www.cosmos.esa.int/gaia), processed by
the Data Processing and Analysis Consortium (DPAC,
https://www.cosmos.esa.int/web/gaia /dpac/consortium) {\it Gaia}.
DPAC support was provided by national institutions, in particular
institutions participating in the multilateral agreement {\it Gaia}.
E.~N. Podzolkova is the recipient of a scholarship from the
Foundation for the Development of Theoretical Physics and Mathematics
"BASIS" (Grant No. 21-2-2-44-1).


\end{document}